\documentclass[9pt,twocolumn,twoside]{optica}

\usepackage{subcaption}
\usepackage{overpic}
\setboolean{shortarticle}{false} 
\usepackage{subeqnar,overpic}
\usepackage{ulem}

\renewcommand{\v}[1]{\ensuremath{\mathbf{#1}}} 
\newcommand{\uv}[1]{\ensuremath{\mathbf{\hat{#1}}}} 
\newcommand{\pd}[2]{\frac{\partial #1}{\partial #2}} 
\newcommand{\pdd}[2]{\frac{\partial^2 #1}{\partial #2^2}}

\ifthenelse{\boolean{shortarticle}}{\colorlet{color2}{color2b}}{\colorlet{color2}{color2}} 

\title{Stable Numerical Schemes for Nonlinear Dispersive Equations with Counter-Propagation and Gain Dynamics}

\author[1]{Chang Sun}
\author[2]{Niall Mangan}
\author[3,6]{Mark Dong}
\author[4]{Herbert G. Winful}
\author[3]{Steven T. Cundiff}
\author[1,5,*]{J. Nathan Kutz}

\affil[1]{Department of Physics, University of Washington, Seattle, WA 90195}
\affil[2]{Department of Engineering Sciences and Applied Mathematics, Northwestern University, Evanston, IL 60208}
\affil[3]{Department of Physics, University of Michigan, Ann Arbor, MI 48109}
\affil[4]{Department of Electrical Engineering and Computer Science, University of Michigan, Ann Arbor, MI 48109}
\affil[5]{Department of Applied Mathematics, University of Washington, Seattle, WA 90195-3925}
\affil[6]{The MITRE Corporation, 202 Burlington Rd. Bedford, MA 01730. The author's affiliation with The MITRE Corporation is provided for identification purposes only and is not intended to convey or imply MITRE's concurrence with, or support for, the positions, opinions, or viewpoints expressed by the author.}
\affil[*]{Corresponding author: sunch610@uw.edu}

\dates{Compiled \today}

\ociscodes{(140.4780)  Optical resonators; (140.3945)   Microcavities;  (060.5530) Pulse propagation and temporal solitons;  (190.0190)  Nonlinear optics}

\begin{abstract}
We develop a stable and efficient numerical scheme for modeling the optical field evolution in a nonlinear dispersive cavity with counter propagating waves and complex, semiconductor physics gain dynamics that are expensive to evalute.
Our stability analysis is characterized by a von-Neumann analysis which shows that many standard numerical schemes are unstable due to competing physical effects in the propagation equations.  
We show that the combination of a predictor-corrector scheme with an operator-splitting not only results in a stable scheme, but provides a highly efficient, single-stage evaluation of the gain dynamics.  
Given that the gain dynamics is the rate-limiting step of the algorithm, our method circumvents the numerical instability induced by the other cavity physics when evaluating the gain in an efficient manner.
We demonstrate the stability and efficiency of the algorithm on a diode laser model which includes three waveguides and semiconductor gain dynamics.  The laser is able to produce a repeating temporal waveform and stable optical comblines, thus demonstrating that frequency combs generation may be possible in chip scale, diode lasers.
\end{abstract}

\setboolean{displaycopyright}{true}

\begin{document}

\maketitle
\thispagestyle{fancy}

\ifthenelse{\boolean{shortarticle}}{\ifthenelse{\boolean{singlecolumn}}{\abscontentformatted}{\abscontent}}{}

\section{Introduction}

Computational methods play a fundamental role in scientific exploration and model development across the physical and engineering sciences.  Simulations help provide critical understanding of physical processes and their interactions in complex systems.  Further, they can provide proof-of-concept engineering designs before expensive manufacturing and/or experiments are performed.  From the aerospace industry to optical laser physics, initial designs are now often test-bedded in simulation environments in order to achieve a qualitative understanding of the physical interactions, good parameter regimes, and robustness of a physical design.   Accurate simulations are also critical for digital twin technologies~\cite{twin}.  Critical in this process is the construction of stable numerical schemes that provides both accuracy and stability for modeling the underlying physics.   In the present work, we develop a robust and stable time-stepping algorithm for modeling counter-propagating waves in a diode laser subject to gain and loss dynamics.  Specifically, we show that an operator splitting scheme with predictor-corrector time-stepping can circumvent the numerical instabilities generated by standard algorithms are that typically used for either counter-propagating waves or for modeling the complex gain dynamics, but not both.  Moreover, the proposed method allows for computational efficiency when modeling the complex semiconductor gain dynamics, thus leading to a robust, stable, and computationally efficient numerical scheme.  The analysis also shows that the scheme can be more broadly applied to nonlinear dispersive wave equations which include competing instability effects from counter-propagating waves and gain dynamics.  

Diode lasers are a ubiquitous technology that have been well developed in theory and experiment.   Emerging research efforts aim to exploit diode lasers in order to generate repeatable waveforms, or mode-locked diode lasers (MLDL), that can be used for robust frequency combs~\cite{MarkIEEE}.   MLDLs can potentially generate frequency combs directly from chip-scale devices \cite{Moskalenko2017, Rosales2011}.   Typically, passive MLDLs comprise two sections: a gain section and a reverse-biased saturable absorber section that leads to the formation of a periodic train of short pulses and hence a comb in the frequency domain. The major obstacle in generating short pulses in diode lasers stems from the nonlinear phase shifts that occur due to fast carrier dynamics \cite{Delfyett1992}, essentially limiting the pulse width inside the cavity. However, single-section diode lasers without saturable absorbers can also operate in a multimode phase-synchronized state known as frequency-modulated (FM) mode locking \cite{Tiemeijer1989}. In the ideal FM mode locked state, the output is a continuous wave in time but the frequency modulation results in a set of comb lines with a fixed, non-zero phase difference. Such FM modelocked operation has been studied most intensively in quantum dot (QD) \cite{Gioannini2015, Rosales2012} and quantum dash \cite{Rosales2012-2} (QDash) lasers, but has also been observed in quantum well (QW) \cite{Sato2003, Calo2015} and bulk semiconductor lasers \cite{Tiemeijer1989}. While some theoretical work has been done for how these combs emerge in a QD single-section laser \cite{Gioannini2015}, only recently has a detailed model for the FM comb generation in the QW diode laser been developed~\cite{MarkIEEE}.  This model builds on various semiconductor quantum well lasers which have varying degrees of complexity. The simplest models include only a single rate equation and photon density variable \cite{Homar1996, Arakawa1986}, while more complex models may use multiple rate equations and more complex forms of the material polarization \cite{KN2010, McDonald1995, Jones1995, Vandermeer2005, Gordon2008, Lenstra2014} with varying degrees of phenomenological expressions and constants inserted.   However, to properly account for how FM combs are in QW lasers, a model must account for the multiple cavity modes as a modulation of the electric field envelope, spectral and spatial hole burning, carrier induced refractive index shift, some intraband carrier dynamics, and cavity dispersion. The gain model builds first principle derivations~\cite{Chow2002, Gioannini2015}, but is tailored to quantum well nanostructures~\cite{MarkIEEE}.

Robust and efficient simulations of the detailed governing equations of the diode laser physics are critical~\cite{MarkIEEE}.  Not only must a time-stepper be developed that is stable, but the complexity of the semiconductor physics requires the implementation of an efficient scheme.  Typically the gain dynamics is modeled with a simple Euler method~\cite{Chow2002, Gioannini2015,MarkIEEE} since this requires only a single stage and can greatly reduce the computational costs.  However, such Euler schemes generate numerical instabilities when the effects of counter-propagating waves and chromatic dispersion are included in the cavity.  Leap-frog schemes, which are at the core of FTDT (finite-difference, time-domain) methods, are quite effective in handling the counter-propagating waves, but they generate numerical instability when modeling the gain dynamics.  Thus the standard schemes produce competing instabilities.  The stability of a numerical scheme is typically evaluated using a von-Neumann analysis~\cite{Kutz:2013}, where conditions for stability, such as constraints on the CFL (Courant-Frederiks-Lewy) number can be explicitly evaluated.  Here we show that an operator splitting technique and predictor-corrector structure allows us to posit a stable and robust scheme for nonlinear, dispersive optical field propagation that is computationally efficient and accurate.  We explicitly demonstrate the instability mechanisms present in the various physics of our model and motivate the development of our stable method.

The paper is outlined as follows:  In Sec.~2, the governing equations for the diode laser physics are presented.   Section~3 evaluates the numerical stability of a variety of time-stepping schemes, showing that the various physical effects in our model produce competing instabilities that require the development of our method.  The numerical stability analysis is performed using a standard von-Neumann analysis.  Section~4 
develops the numerical scheme for the application of the diode laser physics of interest.  Numerical results of our simulation are shown in Sec.~5, with concluding comments in Sec.~6.

\section{Governing Equations}

The diode laser model captures the counter-propagating physics of the electric field $E^{\pm}(z,t)$.   For the specific application where mode-locking is of interest, the standard CW dynamics models must be modified to include chromatic dispersion.  Thus a variety of dominant physical effects are present in the laser:  counter-propagating waves, dispersion, self-phase modulation due to the Kerr effect, dissipation due to cavity losses, and semiconductor gain dynamics.  Aside from the gain dynamics, which is extremely detailed and complex~\cite{Chow2002, Gioannini2015,MarkIEEE}, the remaining physical effects can be easily incorporated into a simple model with distinct interacting terms.    Thus the governing propagation equations for the electric field are given by
\begin{subeqnarray}
&& \hspace*{-.3in} \frac{\partial E^{+}}{\partial z}+\frac{1}{v_{g}}\frac{\partial E^{+}}{\partial t}+i\frac{k''}{2}\frac{\partial^{2}E^{+}}{\partial t^{2}}=-(\alpha_{s}+i\beta_{s})(|E^{+}|^{2}+2|E^{-}|^{2})E^{+} \nonumber \\
&& -i\frac{\omega_{0}}{2c\eta\epsilon_{0}}\Gamma_{xy}P^{+}-\frac{\alpha}{2}E^{+}+S_{sp}^{+}, \\
&&  \hspace*{-.3in} -\frac{\partial E^{-}}{\partial z}+\frac{1}{v_{g}}\frac{\partial E^{-}}{\partial t}+i\frac{k''}{2}\frac{\partial^{2}E^{-}}{\partial t^{2}}=-(\alpha_{s}+i\beta_{s})(|E^{-}|^{2}+2|E^{+}|^{2})E^{-}\nonumber \\
&&-i\frac{\omega_{0}}{2c\eta\epsilon_{0}}\Gamma_{xy}P^{-}-\frac{\alpha}{2}E^{+}+S_{sp}^{-},
\label{eq:full}
\end{subeqnarray}
where $k''\approx1.25ps^{2}/m$ measures the chromatic dispersion in the waveguide~\cite{hudson1,hudson2}, $v_g=c/n_0$ is the group velocity, $\alpha_{s}$ and $\beta_{s}$ are the two-photon absorption and Kerr nonlinear coefficients respectively, $\alpha$ is the linear waveguide loss, and $S_{sp}$ is a spontaneous emission term. The complex, semiconductor gain physics is included in the term  $\Gamma_{xy}$.  This term is discussed in more detail in Sec. 3.  The gain dynamics is computationally expensive to evaluate~\cite{MarkIEEE}, thus a numerical scheme that can efficiently evaluate the $\Gamma_{xy}$ is required.  Specifically, a standard scheme such as 4th-order Runge-Kutta requires a four-stage evaluation process which would require evaluating $\Gamma_{xy}$ four times before updating the solution in the time-stepper.  While the Runge-Kutta scheme has superior stability properties, the four state evaluation would render an already expensive computation significantly longer.  

As will be shown in the next section, the simple structure of the linear terms gives rise to competing numerical instabilities.  Thus a scheme which is ideal for handling counter-propagation, generates an instability due to the presence of the dispersion.  And conversely, a scheme that is well suited for the dispersion easily leads to an instability from the counter-propagation.   These competing instabilities must be circumvented while maintaining a simple evaluation (one-stage) of $\Gamma_{xy}$.

\section{Stability of Time-Stepping Schemes and von-Neumann Analysis}

To simulate the governing equations, appropriate numerical schemes must be considered. In addition to speed and accuracy, the stability of the underlying scheme is of paramount importance. We use standard finite difference discretization schemes and consider their numerical stability properties.   Although an Euler stepping scheme appears to work for the cavity dynamics without dispersion~\cite{MarkIEEE}, the introduction of dispersion guarantees the solution technique is unstable.   To be precise, we consider a subset of the full equations Eqs.~(\ref{eq:full}).  Specifically, we consider the interaction of propagation and dispersion.  Thus the governing linear equations considered are
\begin{equation}
\frac{\partial E}{\partial z}=\frac{1}{v_{g}}\frac{\partial E}{\partial t}+iD\frac{\partial^{2}E}{\partial t^{2}}.
\label{eq:test}
\end{equation}
This partial differential equation thus considers the interaction of one-way wave propagation with speed $v_g$ and chromatic dispersion with strength $D$. A 
von-Neumann analysis can give the stability of different numerical time stepping methods. Specifically,  finite difference discretization of the electric field gives a finite number of discrete spatial and temporal points. A von-Neumann analysis proceeds by letting~\cite{Kutz:2013}
\begin{equation}
E(z_{m},t_{n})=E^{m}_{n}=g^{m}exp(i\xi_{n}h),
\label{eq:von}
\end{equation}
where $\xi_{n}=n\xi$ with $\xi$ defining the numerical step taken in the propagation direction $z$.  Thus the index $n$ denotes the discretization of the field in time, while $m$ denotes the discretization along the propagation direction $z$.      Substituting $E_{n}^{m}$ into the discretized equations associated with Eq.~(\ref{eq:test})  yields an iterative equation for $g$.  Importantly, the norm of $|g|$ determines the stability of the time stepping scheme.   Specifically, if $|g|>1$ the iteration scheme is unstable, while $|g|<1$ gives the potential for a stable scheme since it is linearly stable.   A von-Neumann analysis aims to determine $|g|$ so that stability can be evaluated.

\subsection{Euler time-stepping scheme}
Defining $\Delta z$ and $\Delta t$  as the discretization parameters for $z$ and $t$ respectively, implementation of the Euler time-stepping scheme~\cite{Kutz:2013} gives the discretization of Eq.~(\ref{eq:test}) as
\begin{equation}
\frac{E_{n}^{m+1}-E_{n}^{m}}{\Delta z}=\frac{E_{n+1}^{m}-E_{n-1}^{m}}{2v_{g}\Delta t}+\frac{iD}{\Delta t^{2}}(E_{n+1}^{m}-2E^{m}_{n}+E_{n-1}^{m}),
\end{equation} 
substituting the von-Neumann decomposition Eq.~(\ref{eq:von}) into the above and simplifying yields
%
%
%
\begin{equation}
g=1+i(\frac{v_g\Delta z}{\Delta t}sin\xi h-4\frac{D\Delta z}{\Delta t^{2}}sin^{2}\frac{\xi h}{2}).
\end{equation}
which allow us to explicitly evaluate the stability with
\begin{equation}
|g|^{2}=1+\left(\frac{\Delta z}{v_{g}\Delta t}sin\xi h-4\frac{D\Delta z}{\Delta t^{2}}sin^{2}\frac{\xi h}{2}\right)^{2}>1.
\end{equation}
The von-Neumann analysis ensures that $|g|>1$ for all $\Delta z$ and $\Delta t$ so that the Euler time-stepping scheme in this case is guaranteed to produce numerical instability.  Thus the inclusion of these terms in the full model of Eq.~(\ref{eq:full}) makes an Euler scheme unstable.  Note that without the dispersion and counter-propagation, the Euler method was exactly what was used previously~\cite{Chow2002, Gioannini2015}.

\subsection{Backward Euler time-stepping scheme}
A standard way to consider to stabilize the Euler scheme is implicit formulation~\cite{Kutz:2013}.   Indeed, implicit methods are generally more favorable when the algorithm stability is considered.  However, the implicit formulation comes at a price as will be shown.  Specifically, applying the backward Euler scheme on the wave equation with both propagation and chromatic dispersion, we obtain
\begin{equation}
\frac{E_{n}^{m+1}-E_{n}^{m}}{\Delta z}=\frac{E_{n+1}^{m+1}-E_{n-1}^{m+1}}{2v_{g}\Delta t}+\frac{iD\Delta z}{\Delta t^{2}}(E_{n+1}^{m+1}-2E_{n}^{m+1}+E_{n-1}^{m+1}),
\end{equation} 
%
This can be compared with the standard Euler time-stepping algorithm to see that the future state of the system is required to evaluate the time-step.  For linear equations, this is not necessarily  problematic, but it is often difficult to accomplish for nonlinear schemes.  

Again substituting the von-Neumann decomposition Eq.~(\ref{eq:von}) into the above and simplifying yields
%
%
%
%
%
\begin{equation}
|g|=\frac{1}{\sqrt{1+(\frac{\Delta z}{v_{g}\Delta t}\sin\xi h+\frac{2D\Delta z}{\Delta t^{2}}(\cos\xi h-1))^{2}}}<1.
\end{equation}
Consequently the backward Euler stepping scheme is more robust than the forward Euler scheme since $|g|<1$ is satisfied with no constrains on the discretization size $\Delta z$ and $\Delta t$. In fact, implicit stepping schemes are known to be exceptionally stable but are less used since it's more expensive than the explicit stepping schemes. Specifically, to obtain information of the future steps, the predictor-corrector scheme is consequently used, which can be extremely time-consuming when it's applied to the full wave equation with the existence of the gain term. We are thus interested in developing a less time-consuming but more generally robust stepping scheme. 

\subsection{Leap-frog (2,2) time-stepping scheme}
The discretization form of the equation including propagation and dispersion with a leap-frog (2,2) method~\cite{Kutz:2013} is given by
\begin{equation}
\frac{E_{n}^{m+1}-E_{n}^{m-1}}{2\Delta z}=\frac{E_{n+1}^{m}-E_{n-1}^{m}}{2v_{g}\Delta t}+\frac{iD}{\Delta t^{2}}(E_{n+1}^{m}-2E^{m}_{n}+E_{n-1}^{m}).
\end{equation} 
Again substituting the von-Neumann decomposition Eq.~(\ref{eq:von}) into the above and simplifying yields
%
%
%
%
%
\begin{equation}
g=\frac{1}{2}(iM\pm\sqrt{4-M^{2}}).
\end{equation}
where
\begin{equation}
M =2\frac{\Delta z}{v_{g}\Delta t} \sin\xi h+4\frac{D\Delta z}{\Delta t^{2}}(\cos\xi h-1).
\end{equation}

Depending on whether $4-M^{2}\geq0$, we have
\begin{equation}
g=\left\{
\begin{array}{r@{\;\;}l}
\pm\frac{1}{2}\sqrt{4-M^{2}}+i\frac{M}{2}, & 4-M^{2}\geq0 ,\\
\frac{i}{2}(M\pm\sqrt{M^{2}-4}), & 4-M^{2}<0,\\
\end{array}
\right.
\end{equation}
For $4-M^{2}<0$, 
\begin{equation}
|g|^{2}=\frac{1}{4}(M\pm\sqrt{M^{2}-4})^{2}=\frac{1}{2}(M^{2}-2\pm M\sqrt{M^{2}-4}).
\end{equation} 
In this case $M^{2}>4$ which gives 
\begin{equation}
|g|^{2}=\frac{1}{2}(M^{2}-2+ M\sqrt{M^{2}-4}),
\end{equation} 
so that $|g|^{2}>1$ and the leap-frog (2,2) time-stepping scheme is unstable.

\begin{figure}[t]
{\begin{overpic}
[width=0.95\linewidth]{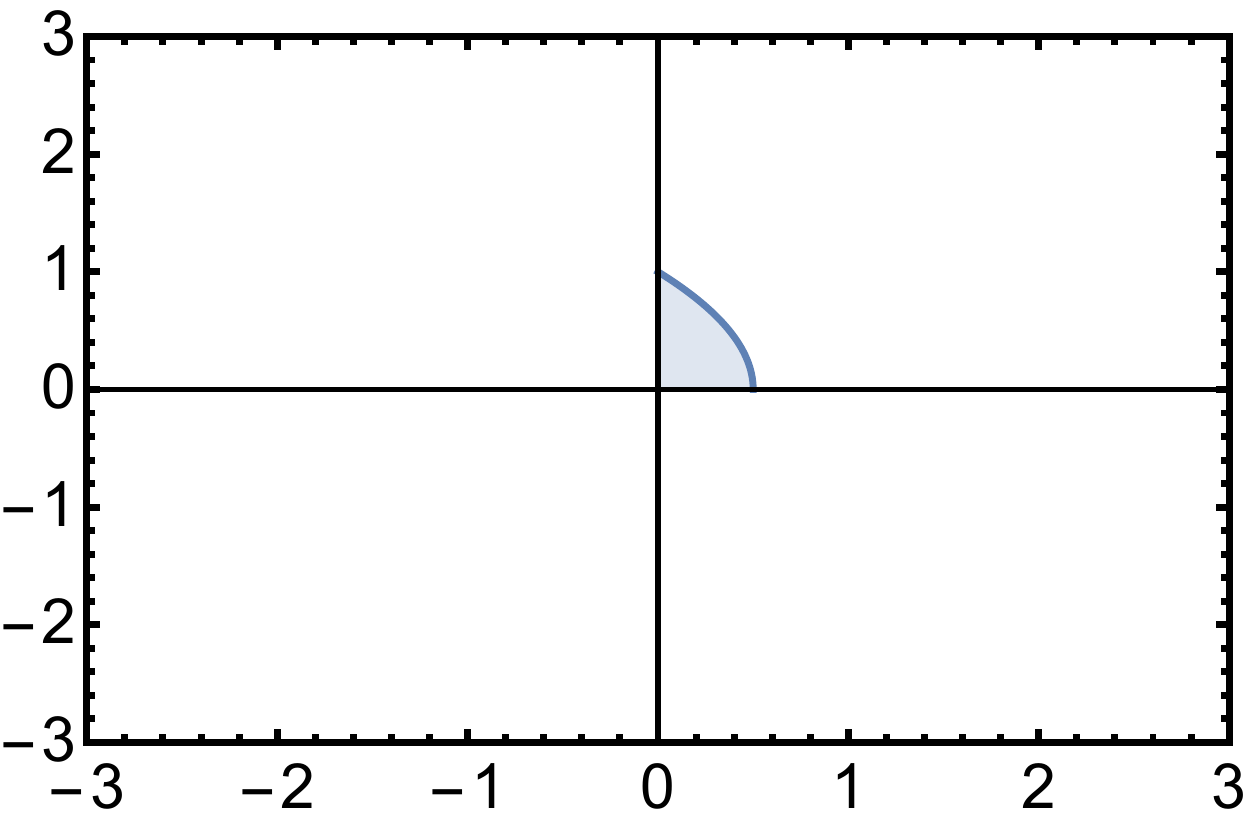}
\put(-5,29){{\large $\frac{\Delta z}{V_g \Delta t}$ }}
\put(50,-3){{\large $\frac{2D\Delta z}{\Delta t^2}$ }}
\put(73,20.5){\large unstable}
\put(59,40){\large stable}
\end{overpic}}
\caption{Regions for stable stepping of the wave equation with propagation and chromatic dispersion using the leap-frog (2,2) scheme.}
\label{StabilityRegion}
\end{figure}

For $4-M^{2}\geq0$, 
\begin{equation}
|g|^{2}=\frac{1}{4}(4-M^{2})+\frac{M^{2}}{4}=1,
\end{equation}
thus the algorithm is at the stability boundary  $|g|=1$.
For this case, we obtain the constraints on $M$ as $-2\leq M\leq 2$, that is, 
\begin{equation}
 -1\leq \frac{\Delta z}{v_{g}\Delta t}sin\xi h+2\frac{D\Delta z}{\Delta t^{2}}(cos\xi h-1)\leq 1.
\end{equation}
This constraint can be satisfied by selecting the discretization sizes $\Delta z$ and $\Delta t$ as shown in Fig. ~\ref{StabilityRegion}. However, the inclusion of the gain terms into the leap-frog (2,2) scheme generates instability, thus we need a more robust numerical scheme for the more general case. 

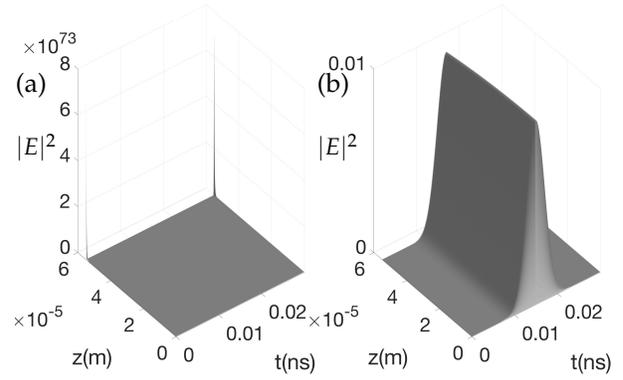
\begin{figure}[t]
{\begin{overpic}
[width=0.95\linewidth]{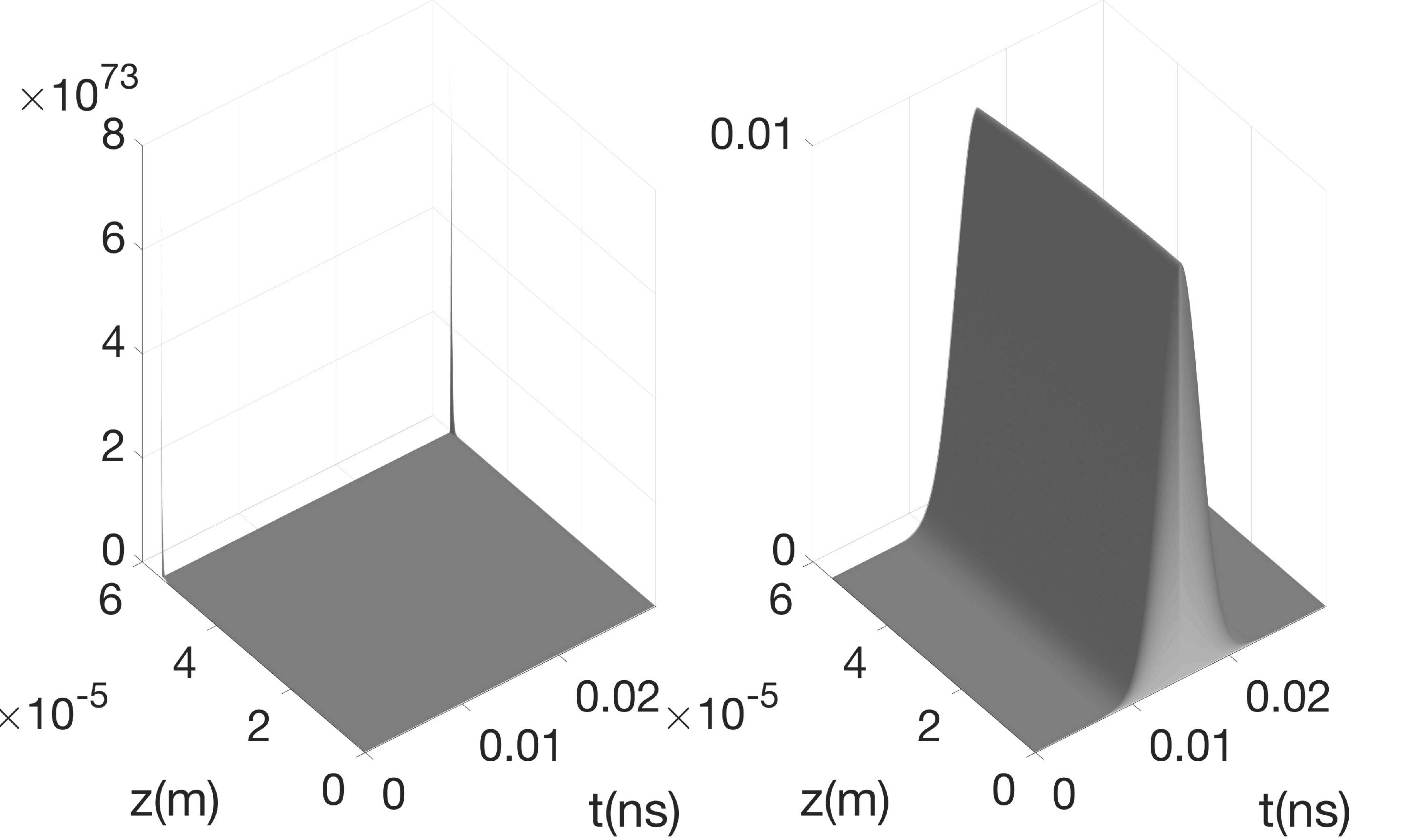}
\put(0,45){{\large (a) }}
\put(48,45){{\large (b) }}
\put(0,35){{$|E|^2$}}
\put(48,35){{$|E|^2$}}
\end{overpic}}
\caption{Associated numerical results of the unstable and stable region using leap frog (2,2). (a) The numerical solution quickly blows up without satisfying the constraints on the discretized steps $\Delta z$ and $\Delta t$. (b) Numerical result consistently demonstrates the stability within the shaded region in Fig. ~\ref{StabilityRegion}.}
\label{Unstable}
\end{figure}

\subsection{Predictor-corrector time-stepping scheme}
\begin{figure}[t]
{\begin{overpic}
[width=0.95\linewidth]{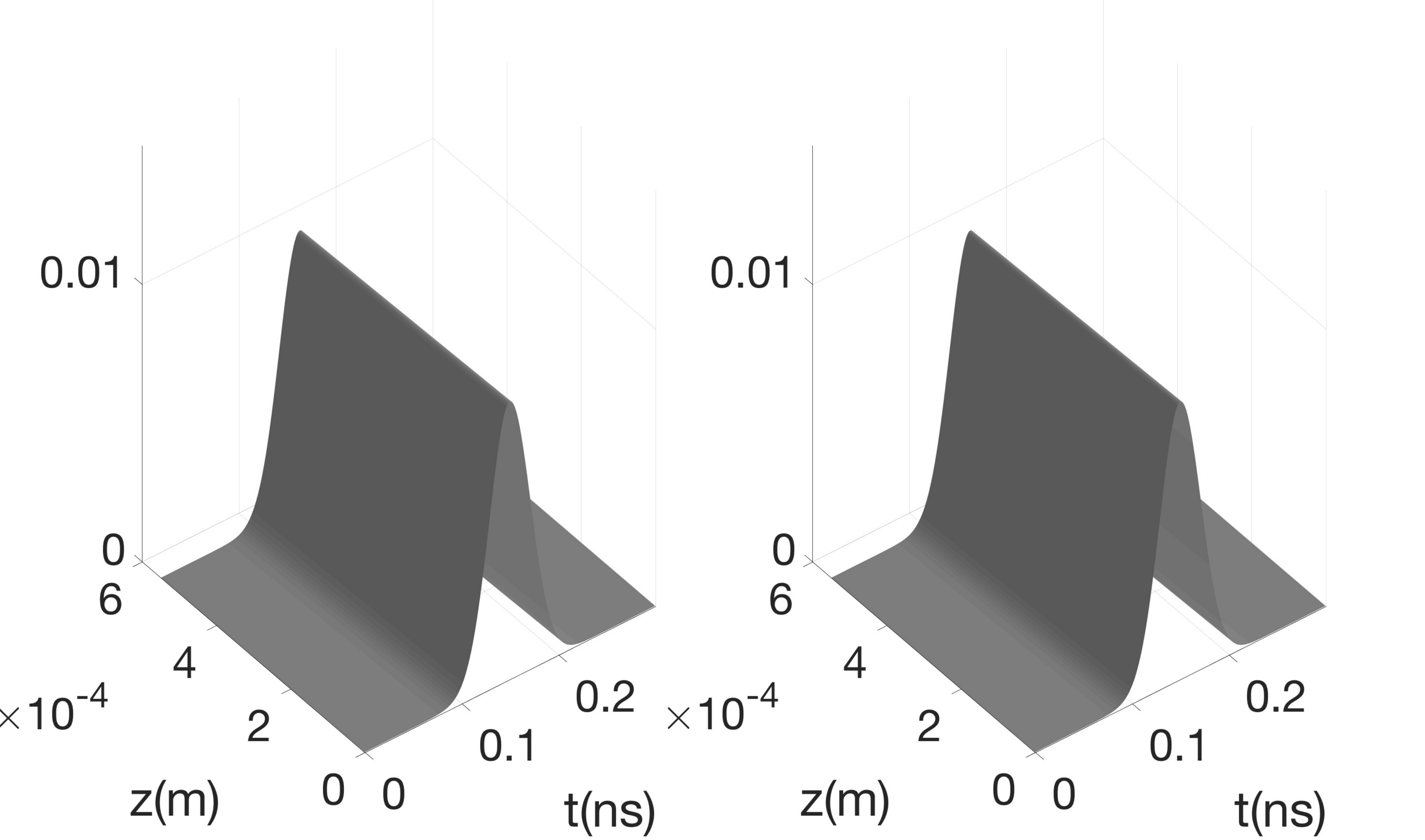}
\put(0,45){{\large (a) }}
\put(48,45){{\large (b) }}
\put(0,32){{$|E|^2$}}
\put(48,32){{$|E|^2$}}
\end{overpic}}
\caption{Numerical results of the wave equation with propagation and dispersion. (a) Predictor-corrector efficiently prevents the solution from explosion without constraints on the size of discretization. (b) Numerical simulation using FFT shows consistency with the predictor corrector scheme.}
\label{PC}
\end{figure}
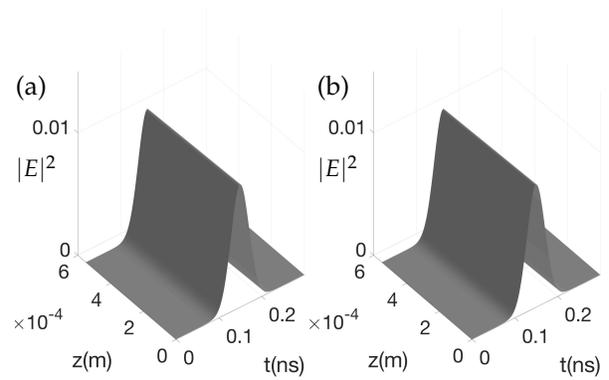

Considering the more general case that includes propagation, chromatic dispersion and gain, the predictor-corrector scheme provides an appropriate stable implementation which makes use of both forward and backward Euler time-stepping due to their speed and stability respectively. Specifically, we consider the more general equation
\begin{equation}
\frac{\partial E}{\partial z}=\frac{1}{v_g}\frac{\partial E}{\partial t}+iD\frac{\partial^{2}E}{\partial t^{2}}+f(E),
\end{equation}
where gain terms are now included as the function $f(E)$. Euler time-stepping of the full equations is used in the predictor step and gives the propagated electric field as
\begin{equation}
E_{n}^{pred}\!\!=\!\!E_{n}^{m}\!+\! \Delta z  \!\! \left[\!\! \frac{E_{n+1}^{m}\!-\!E_{n-1}^{m}}{2v_{g}\Delta t}\\
+\frac{iD}{\Delta t^{2}}(E_{n+1}^{m}\!-\!2E^{m}_{n}\!+\!E_{n-1}^{m})\!+\!f(E_{n}^{m}) \!\right] \!\!.
\end{equation}
Substituting $E_{n}^{pred}$ into the backward Euler time-stepping scheme gives 
\begin{equation}
E_{n}^{m+1}\!\!=\!\!E_{n}^{m}\!+ \!\Delta z \!\!\! \left[ \! \!\frac{E_{n+1}^{pred}\!-\!E_{n-1}^{pred}}{2v_{g}\Delta t}
\!+\! \frac{iD}{\Delta t^{2}}(\!E_{n+1}^{pred}\!-\!2E_{n}^{pred}\!+\!\!E_{n-1}^{pred})\!+\!\!f(E_{n}^{m}) \! \!\right] \!\!.
\end{equation} 
This numerical scheme is stable without constraints on the discretized steps of $\Delta z$ and $\Delta t$, as shown in Fig. ~\ref{PC}. The equation with both propagation and dispersion is evaluated with the predictor-corrector scheme along the characteristic of $\frac{\Delta z}{\Delta t}=v_g$, and is compared to the case in which the identical equation is evaluated by the spectral method FFT~\cite{Kutz:2013}. The consistency shows the reliability of this predictor-corrector scheme. Note that the numerical discretized scheme works with non-periodic boundary conditions, while the spectral schemes are only capable of dealing with specific types of boundary conditions.  For modeling of diode lasers, the boundary conditions do not allow us to use spectral methods.

This time-stepping scheme gives a robust numerical implementation for the full wave equation including the gain term, propagation and dispersion. However, the algorithm also forces one to evaluate the gain term two time per numerical step:  once for the prediction step, and once for the corrector step.  But the evaluation of the semiconductor gain dynamics is extremely expensive and time-consuming~\cite{Chow2002, Gioannini2015,MarkIEEE}.  Indeed, it is highly preferable to develop a robust and stable scheme that only evaluates the gain dynamics once per time step.

\subsection{Operator splitting techniques}
Inspired from the knowledge that linearly coupled physical effects can be decoupled from each other, the partial differential equation of propagating, dispersion and complex gain terms can be discretized separately including only individual effects from each of the these terms~\cite{Kutz:2013}. Specifically, the wave propagation and dispersion act independently (approximately) of the gain over short time steps, and vice versa for the gain. By defining the gain term to be $f(E)$, we can consequently split the equation into two pieces as below:
\begin{equation}
\frac{\partial E}{\partial z}=c\frac{\partial E}{\partial t}+iD\frac{\partial^{2}E}{\partial t^{2}},
\end{equation}
and
\begin{equation}
\frac{\partial E}{\partial z}=f(E).
\end{equation}
The first equation is on the stability boundary of the leap-frog (2,2) scheme, while the second equation is known to be stably solved with Euler time stepping as already demonstrated in previous work~\cite{Chow2002, Gioannini2015,MarkIEEE}. This numerical scheme with the operator-splitting technique implemented is stable and efficient for solving the full wave equation with all terms included.  Note that leap-frog (2,2) is only capable of pushing $g$ to the stability boundary of $|g|=1$ with $M^{2}\leq4$ satisfied, while $|g|<1$ can be satisfied without constraints using the backward Euler scheme.  A new time stepping method making use of both operator-splitting techniques and predictor-corrector is preferable. Specifically, we make use of the Euler method for the propagation of the gain term while applying the predictor-corrector to the rest of the equation which includes only propagation and chromatic dispersion:
\begin{equation}
E_{n}^{pred}\!=\!E_{n}^{m}\!+\!\frac{\Delta z(E_{n+1}^{m}-E_{n-1}^{m})}{2v_{g}\Delta t}+\frac{iD\Delta z}{\Delta t^{2}}(E_{n+1}^{m}-2E^{m}_{n}+E_{n-1}^{m}),
\end{equation}
and 
\begin{equation}
E_{n}^{m+1}\!=\!E_{n}^{m}\!+\!\frac{\Delta z}{2v_{g}\Delta t}(E_{n+1}^{pred}\!-\!E_{n-1}^{pred})\!+\!\frac{iD\Delta z}{\Delta t^{2}}(E_{n+1}^{pred}\!-\!2E_{n}^{pred}\!+\!E_{n-1}^{pred}).
\end{equation} 
By taking the combination of both predictor-corrector and operator-splitting, we are capable of evaluating the gain term more effectively (at half the computational cost) while maintaining the stability of the entire numerical scheme. 

\section{Numerical scheme for the laser cavity}
To stably and efficiently incorporate the dispersion, wave propagation and gain into the numerical scheme, we employ the advocated combination of predictor-corrector and operator splitting methods.
To simplify the full governing Eqs.~(\ref{eq:full}),  the semiconductor population dynamics are grouped into $f_{p}$ and $f_{b}$:
\begin{equation}
\begin{split}
f_{p}(E^{+},E^{-})=-(\alpha_{s}+i\beta_{s})(|E^{+}|^{2}+2|E^{-}|^{2})E^{+}\\
-i\frac{\omega_{0}}{2c\eta\epsilon_{0}}\Gamma_{xy}P^{+}-\frac{\alpha}{2}E^{+}+S_{sp}^{+},
\end{split}
\end{equation}
\begin{equation}
\begin{split}
f_{b}(E^{+},E^{-})=-(\alpha_{s}+i\beta_{s})(|E^{-}|^{2}+2|E^{+}|^{2})E^{-}\\
-i\frac{\omega_{0}}{2c\eta\epsilon_{0}}\Gamma_{xy}P^{-}-\frac{\alpha}{2}E^{+}+S_{sp}^{-}.
\end{split}
\end{equation}
Making this substitutions into the governing equations and rearranging we construct our system of equations
\begin{equation}
\frac{\partial E^{+}}{\partial z}=-\frac{1}{v_{g}}\frac{\partial E^{+}}{\partial t}-i\frac{k''}{2}\frac{\partial^{2}E^{+}}{\partial t^{2}}+f_{p}(E^+, E^-),
\end{equation}
\begin{equation}
-\frac{\partial E^{-}}{\partial z}=-\frac{1}{v_{g}}\frac{\partial E^{-}}{\partial t}-i\frac{k''}{2}\frac{\partial^{2}E^{-}}{\partial t^{2}}+f_{b}(E^+, E^-).
\end{equation}

\subsection{Operator-splitting technique and predictor-corrector scheme}
We separate the dispersion and propagation from the gain dynamics  so that the it can  be discretized using a simple forward Euler scheme.  Thus we can
%
propagate along the $z$ direction
\begin{equation}
E^{+}_{m+1,n}=E^{+}_{m,n}+\Delta zf_{p}(E^{+}_{m,n},E^{-}_{m,n}),
\end{equation}
\begin{equation}
E^{-}_{m+1,n}=E^{-}_{m,n}-\Delta zf_{b}(E^{+}_{m,n},E^{-}_{m,n}).
\end{equation}

We now consider the remaining terms in the governing equation which  include the propagation and dispersion, 
\begin{equation}
\frac{\partial E^{+}}{\partial z}=-\frac{1}{v_{g}}\frac{\partial E^{+}}{\partial t}-i\frac{k''}{2}\frac{\partial^{2}E^{+}}{\partial t^{2}},
\end{equation}
\begin{equation}
-\frac{\partial E^{-}}{\partial z}=-\frac{1}{v_{g}}\frac{\partial E^{-}}{\partial t}-i\frac{k''}{2}\frac{\partial^{2}E^{-}}{\partial t^{2}}.
\end{equation}
Leap-frog (2,2) is used as the predictor step:
\begin{equation}
\begin{split}
\frac{E_{m+1,n}^{+}-E_{m-1,n}^{+}}{2\Delta z}=-\frac{1}{2v_{g}\Delta t}(E_{m,n+1}^{+}-E_{m,n-1}^{+})\\
+\frac{iD}{\Delta t^{2}}(E_{m,n+1}^{+}-2E^{+}_{m,n}+E_{m,n-1}^{+}),
\end{split}
\end{equation} 
\begin{equation}
\begin{split}
\frac{E_{m-1,n}^{-}-E_{m+1,n}^{-}}{2\Delta z}=-\frac{1}{2v_{g}\Delta t}(E_{m,n+1}^{-}-E_{m,n-1}^{-})\\
+\frac{iD}{\Delta t^{2}}(E_{m,n+1}^{-}-2E^{-}_{m,n}+E_{m,n-1}^{-}),
\end{split}
\end{equation} 
from which we obtain the predicted $E^{\pm}$ as  
\begin{equation}
\begin{split}
E_{pre,n}^{+}=E_{m-1,n}^{+}+\frac{\Delta z}{v_{g}\Delta t}(E_{m,n-1}^{+}-E_{m,n+1}^{+})\\
+\frac{2iD\Delta z}{\Delta t^{2}}(E_{m,n+1}^{+}-2E^{+}_{m,n}+E_{m,n-1}^{+}),
\end{split}
\end{equation} 
\begin{equation}
\begin{split}
E_{pre,n}^{-}=E_{m-1,n}^{-}-\frac{\Delta z}{v_{g}\Delta t}(E_{m,n-1}^{-}-E_{m,n+1}^{-})\\
-\frac{2iD\Delta z}{\Delta t^{2}}(E_{m,n+1}^{-}-2E^{-}_{m,n}+E_{m,n-1}^{-}).
\end{split}
\end{equation} 

Substituting $E_{pre,n}^{+}$ and $E_{pre,n}^{-}$ into the backward Euler scheme we obtain
\begin{equation}
\begin{split}
E_{m+1,n}^{+}=E_{m,n}^{+}+\frac{\Delta z}{2v_{g}\Delta t}(E_{pre,n-1}^{+}-E_{pre, n+1}^{+})\\
+\frac{iD\Delta z}{\Delta t^{2}}(E_{pre,n+1}^{+}-2E_{pre,n}^{+}+E_{pre,n-1}^{+}),
\end{split}
\end{equation}
\begin{equation}
\begin{split}
E_{m+1,n}^{-}=E_{m,n}^{-}-\frac{\Delta z}{2v_{g}\Delta t}(E_{pre,n-1}^{-}-E_{pre, n+1}^{-})\\
-\frac{iD\Delta z}{\Delta t^{2}}(E_{pre,n+1}^{-}-2E_{pre,n}^{-}+E_{pre,n-1}^{-}).
\end{split}
\end{equation}
Note that for the last two time steps, $E^{\pm}_{m,n}$ and $E^{\pm}_{m,n-1}$ must both be saved.  Further note that leap-frog (2,2) can be employed directly without the corrector step if the discretization sizes $\Delta z$ and $\Delta t$ are elaborately selected for $|g|$ to be on the stability boundary.  But with the backward Euler stepping as a corrector, we are capable of achieving a more robust time-stepping algorithm without constraints on $\Delta z$ and $\Delta t$.

\subsection{Propagation along the direction of time}

The algorithm presented in the previous section advanced the solution along the propagation direction $z$ in increments of $\Delta z$.  We can alternatively advance the solution in $t$ in increments of $\Delta t$.  In this case, we re-arrange the governing equations as follows:
%
%
\begin{equation}
\frac{1}{v_{g}}\frac{\partial E^{+}}{\partial t}=-\frac{\partial E^{+}}{\partial z}-i\frac{k''}{2}\frac{\partial^{2}E^{+}}{\partial t^{2}}+f_{p}(E^+, E^-),
\end{equation}
\begin{equation}
\frac{1}{v_{g}}\frac{\partial E^{-}}{\partial t}=\frac{\partial E^{-}}{\partial z}-i\frac{k''}{2}\frac{\partial^{2}E^{-}}{\partial t^{2}}+f_{b}(E^+, E^-).
\end{equation}
We then can time-step by evaluating the electric field time derivatives along characteristics. This is equivalent to discretizing each time and spatial step so that $\frac{\Delta z}{\Delta t}=v_g$. This is highly stable for the traveling wave equations, allowing large time steps. Neglecting the dispersion term for the moment, and discretizing the rest of the equations we obtain
\begin{equation}
\frac{E^{+}_{j+1,n+1}-E^{+}_{j+1,n}}{\Delta t}=v_g(-\frac{E^{+}_{j+1,n}-E^{+}_{j,n}}{\Delta z}+f_{p}(E^{+},E^{-})),
\end{equation}
\begin{equation}
\frac{E^{-}_{j,n+1}-E^{-}_{j,n}}{\Delta t}=v_g(\frac{E^{-}_{j+1,n}-E^{-}_{j,n}}{\Delta z}+f_{b}(E^{+},E^{-})).
\end{equation}
The indices $j$ and $n$, indicate spatial discretization and temporal discretization respectively. Note that
the temporal discretization for the forward wave, $E^{+}$, is at spatial index $j$ while the backward wave, $E^-$ is at $j+1$. Rearranging the equations
\begin{equation}
E^{+}_{j+1,n+1}=E^{+}_{j,n}+\Delta zf_{p}(E^{+},E^{-}),
\end{equation}
\begin{equation}
E^{-}_{j,n+1}=E^{-}_{j+1,n}+\Delta zf_{b}(E^{+},E^{-}).
\end{equation}
To incorporate the dispersion term, we use a predictor-corrector method in the time-stepping approximation for the
dispersion using a second-order accurate scheme,
\begin{equation}
\frac{\partial^2 E_{j,n}^{\pm}}{\partial t^{2}}=\frac{E^{\pm}_{j,n+1}-2E^{\pm}_{j,n}+E^{\pm}_{j,n-1}}{\Delta t^2}.
\end{equation}

The predictor step calculates $E^{\pm}$ without dispersion as
\begin{equation}
E^{+}_{j+1,pred}=E^{+}_{j,n}+\Delta zf_{p}(E^{+}_{j,n},E^{-}_{j,n}),
\end{equation}
\begin{equation}
E^{-}_{j,pred}=E^{-}_{j+1,n}+\Delta zf_{b}(E^{+}_{j+1,n},E^{-}_{j+1,n}).
\end{equation}
Then the corrector step calculates $E^{\pm}$ with dispersion, using the predicted values as
\begin{equation}
\begin{split}
E^{+}_{j+1,n+1}=E^{+}_{j,n}-i\frac{k''\Delta z}{2}\frac{E^{+}_{j,pred}-2E^{+}_{j,n}+E^{+}_{j,n-1}}{\Delta t^2}\\
+\Delta zf_{p}(E^{+}_{j,n},E^{-}_{j,n}),
\end{split}
\end{equation}
\begin{equation}
\begin{split}
E^{-}_{j,n+1}=E^{-}_{j+1,n}-i\frac{k''\Delta z}{2}\frac{E^{-}_{j+1,pred}-2E^{-}_{j+1,n}+E^{-}_{j+1,n-1}}{\Delta t^2}\\
+\Delta zf_{b}(E^{+}_{j+1,n},E^{-}_{j+1,n}).
\end{split}
\end{equation}
Note that in the last two time steps, it is also required that $E^{\pm}_{j,n-1}$ and $E^{\pm}_{j,n}$ must both be saved. This algorithm can be further accelerated when combined with the operator-splitting technique. Specifically, we can propagate the gain terms $f_{p}$ and $f_{b}$ separately using Euler stepping, whereas the chromatic dispersion and wave propagation are evaluated using the predictor-corrector.   We can consequently obtain a second robust and efficient numerical method of the full wave equation. 

Notice that this algorithm is capable of numerically evaluating the wave equations with unconvential boundary conditions. Specifically, the dynamics inside a cavity may require the reflecting boundary conditions  
\begin{equation}
\begin{split}
E^{+}_{0,n+1}=-rE^{-}_{0,n},\\
E^{-}_{L,n+1}=-rE^{+}_{L,n},
\end{split}
\end{equation}
where $L$ is the left most spatial index, and $r$ is the reflection coefficient. In more general cases, a filter may be implemented. That is, the electric field at the edge in frequency domain is multiplied by a Lorentzian:
\begin{equation}
\tilde{E}^{-}(z=L,\omega)=r\tilde{E}^{+}(z=L,\omega)\mathbf{L}(\omega),
\end{equation}
\begin{equation}
\mathbf{L}(\omega)=\frac{\Gamma}{i(\omega_{0}-\omega)-\Gamma}.
\end{equation}
The filter function is dimensionless and has a maximum of $\mathbf{L}(\omega=\omega_{0})=1$. Transforming this back into time domain, we have the expression as 
\begin{equation}
E^{-}(z=L,t)=r\Gamma\int_{-\infty}^{t}dt'e^{i\omega_{0}(t-t')-\Gamma(t-t')}E^{+}(z=L,t').
\end{equation}
Spectral methods are not appropriate for implementation due to these boundary conditions. Specifically, Fourier based methods require periodic boundary conditions while Chebychev polynomials assume either Dirichlet or Neumann condition be imposed~\cite{Kutz:2013}. The filter on one side of the cavity and the reflector on the other side rule out these fast spectral methods. In contrast, the standard finite difference discretization schemes without any assumption on the boundary conditions are still appropriate for implementation, where the boundary filter integral can be discretized in time and the resulting sum performed at each time step of the simulation.

\section{Numerical Simulations}
For our simulations of the complete governing equations, we used the gain model for a quantum well based semiconductor laser by Dong et al.~\cite{MarkIEEE}. We proceed by giving the essential equations to be solved, leaving the detailed derivation of the governing equations for the Appendix.  The total electric field in the cavity is taken as a sum of forward and backward propagating components
\begin{align}
E(z,t) = E_+(z,t) e^{ik_0 z}+E_-(z,t)e^{-i k_0 z}
\end{align}
whose amplitudes satisfy the slowly-varying envelope equation
\begin{align}
\label{tw_eqn}
\pm\pd{}{z}E_\pm(z,t) + \frac{1}{v_g} \pd{}{t}E_\pm(z,t) = \Gamma_{xy} \frac{\omega_0^2}{2i k_0 c^2 \epsilon_0} \langle P_{tot}(t) e^{\mp ik_0z} \rangle
\end{align}
where the angular brackets signify averaging over a few wavelengths.  Here, $v_g = c/n_0$ is the group velocity, $n_0$ is the group refractive index, $\Gamma_{xy}$ is the transverse confinement factor, $\omega_0$ is the central photon frequency (the choice of $\omega_0$ can be arbitrary but is generally chosen to be the transition frequency at the band edge), and $k_0 = n_0 \omega_0/c$. With Bloch equations tailored to semiconductors as well as the standard adiabatic approximation (see Appendix), the total polarization for a 2-D quantum well can be written as:
\begin{align}
\label{ptot_sumk2}
P_{tot}(t) = \frac{2}{V} \sum_{\v{k}} d^*_{ev} p(\v{k},t) = i\frac{|d_{cv}|^2}{2\hbar\Gamma} \frac{2}{V} \sum_{\v{k}} (\rho^e_{E_t}+\rho^h_{E_t}-1) F(E_t,z,t).
\end{align}

With a simple parabolic dispersion relation and converting the $\v{k}$-summation to a transverse energy integral, we obtain:
\begin{align}
\label{ptot_t_main}
P_{tot}(t) =  i\frac{|d_{cv}|^2}{2\hbar\Gamma} \int dE_t D_r^{2D} (\rho^e_{E_t}+\rho^h_{E_t}-1) F(E_t,z,t)
\end{align}
The dipole matrix element can be rewritten as the momentum matrix element via $|d_{cv}|^2 = \frac{q^2}{m_0^2 \omega_0^2} |\uv{e}\cdot \v{p}|^2$ where $q$ is the electron charge and $m_0$ the electron mass. The macroscopic polarization calculated in Eq.~(\ref{ptot_t_main}) serves as a source term for the forward and backward propagating electric fields in the laser.  The constants on the right hand side of \ref{tw_eqn} can be combined to yield a gain coefficient 
\begin{align*}
g_0 &=  \frac{\Gamma_{xy}q^2 D^{2D}_r |\uv{e}_j \cdot \v{p}_{cv}|^2}{2 n_0 c \epsilon_0 m_0^2\Gamma}.
\end{align*}

To complete the derivation of the propagation equations, we include the effects of carrier gratings resulting from the interference between forward and backward waves. Our approach to modeling this spatial hole burning (SHB) is to follow the techniques of \cite{Homar1996}, \cite{Javaloyes2009} and \cite{Homar1996-2} and expand the QW population into its second harmonic in space. In this formulation, the population becomes
\begin{align}
\label{p_zexp}
\rho^{e,h}_{E_t} = \rho^{e,h}_{qw, E_t} + \rho_{g,E_t} e^{i2k_0z} + \rho^*_{g,E_t} e^{-i2k_0z} + ...
\end{align}
For simplicity, we have used a single variable for the carrier gratings for both electrons and holes. The filtered  field in the polarization also consists of forward and backward components:
\begin{align}
\label{F_zexp}
F &= F_+ e^{-ik_0 z} + F_- e^{ik_0 z}
\end{align}
Inserting Eqs.~(\ref{ptot_t_main}), (\ref{p_zexp}), and  (\ref{F_zexp}) in Eq.~(\ref{tw_eqn}) and keeping only the phase-matched terms we obtain the electric field equations:
\begin{align}
\label{wave_eq_simple}
\begin{split}
\pm\pd{E_\pm}{z}+ &\frac{1}{v_g} \pd{E_\pm}{t} = \frac{g_0}{2} \int \frac{dE_t}{\hbar\omega_0} (\rho^e_{qw, E_t}+\rho^h_{qw,E_t}-1) F_\pm(E_t, z,t)\\
 & + g_0 \int \frac{dE_t}{\hbar\omega_0} \rho^{(*)}_{g,E_t} F_\mp(E_t, z,t)
\end{split}
\end{align}
We note that the grating term $\rho_{g,E_t}^{(*)}$ is associated with the forward wave equation and its conjugate with the backward wave. 

Finally, we simply add the additional terms in Eq.~(\ref{wave_eq_simple}) that describe standard linear and nonlinear effects, and scale via $n_{qw}$, the number of quantum wells to obtain                                                                                                                                 

\begin{align}
\begin{split}
\label{wave_eq}
\pm \pd{E_\pm}{z}+ &\frac{1}{v_g} \pd{E_\pm}{t} + i\frac{k''}{2} \pdd{E_\pm}{t} = \\
& -\frac{\alpha}{2} E_\pm - \left(\frac{\alpha_S}{2}+i\beta_S\right)(|E_\pm|^2+2|E_\mp|^2)E_\pm +S_{sp} \\
 & + n_{qw} \frac{g_0}{2} \int \frac{dE_t}{\hbar\omega_0} (\rho^e_{qw, E_t}+\rho^h_{qw,E_t}-1) F_\pm(E_t, z,t)\\
 & + n_{qw} g_0 \int \frac{dE_t}{\hbar\omega_0} \rho^{(*)}_{g,E_t} F_\mp(E_t, z,t)
\end{split}
\end{align}
where $k''$ is the dispersion coefficient, $\alpha$ is the linear waveguide loss, and $\alpha_S, \beta_S$ are respectively the two-photon absorption and Kerr nonlinear coefficients, and $S_{sp}$ is the spontaneous emission term derived in the \cite{MarkIEEE}.

For simplicity, rewrite the complicated gain term as a function of $G_{\pm}(E_\pm)$ the for Eq. \ref{wave_eq} we obtain
\begin{align}
\begin{split}
\label{wave_eq_simple2}
\pm \pd{E_\pm}{z}+ &\frac{1}{v_g} \pd{E_\pm}{t} + i\frac{k''}{2} \pdd{E_\pm}{t} = \\
& -\frac{\alpha}{2} E_\pm - \left(\frac{\alpha_S}{2}+i\beta_S\right)(|E_\pm|^2+2|E_\mp|^2)E_\pm \\
& +S_{sp}+G_{\pm}(E_\pm).
\end{split}
\end{align}
These field equations are coupled with the carrier rate equations for the SCH and QW sections, of which the complete forms are shown in the Appendix. 

Unlike the simulations presented in~\cite{MarkIEEE}, we modify the physical model to include multiple waveguides for enhancing nonlinear pulse shaping and mode-locking~\cite{WilliamsWGA}.  Thus the simulations presented here provide potentially new physics for aiding in mode-locking.  
Specifically, we extend the diode model to a waveguide array with three waveguides. By coupling out
low-intensity components to the neighboring waveguides, we can effectively shape the electric field propagating in the first waveguide making use of its effects of intensity discrimination, thus achieve highly performed mode-locked pulses in the laser cavity~\cite{proctor05,kutz08,ching12}.   For nearest neighbor coupling, the resulting approximate evolution dynamics describing the waveguide array mode-locking is given by

\begin{align}
\begin{split}
\label{wave_eq1}
\pm \pd{E^{1}_\pm}{z}+ &\frac{1}{v_g} \pd{E^{1}_\pm}{t} + i\frac{k''}{2} \pdd{E^{1}_\pm}{t} =  \\
 & -\frac{\alpha}{2} E^{1}_\pm - \left(\frac{\alpha_S}{2}+i\beta_S\right)(|E^{1}_\pm|^2+2|E^{1}_\mp|^2)E^{1}_\pm \\
 & +S_{sp} +G^{1}_{\pm}+iCE^{2}_\pm
\end{split}
\end{align}

\begin{align}
\begin{split}
\label{wave_eq2}
\pm \pd{E^{2}_\pm}{z}+ &\frac{1}{v_g} \pd{E^{2}_\pm}{t} + i\frac{k''}{2} \pdd{E^{2}_\pm}{t} =  \\
 & -\frac{\alpha}{2} E^{2}_\pm - \left(\frac{\alpha_S}{2}+i\beta_S\right)(|E^{2}_\pm|^2+2|E^{2}_\mp|^2)E^{2}_\pm \\
 & +S_{sp} +G^{2}_{\pm}+iC(E^{1}_\pm+E^{3}_\pm)
\end{split}
\end{align}

\begin{align}
\begin{split}
\label{wave_eq3}
\pm \pd{E^{3}_\pm}{z}+ &\frac{1}{v_g} \pd{E^{3}_\pm}{t} + i\frac{k''}{2} \pdd{E^{3}_\pm}{t} =  \\
 & -\frac{\alpha}{2} E^{3}_\pm - \left(\frac{\alpha_S}{2}+i\beta_S\right)(|E^{3}_\pm|^2+2|E^{3}_\mp|^2)E^{3}_\pm \\
 & +S_{sp} +G^{3}_{\pm} +iCE^{2}_\pm
\end{split}
\end{align}

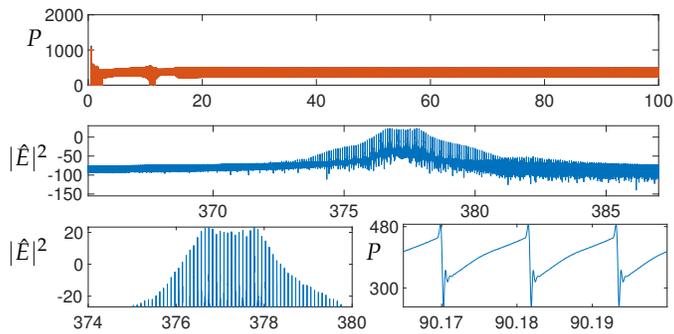
\begin{figure}[t]
{\begin{overpic}
[width=0.95\linewidth]{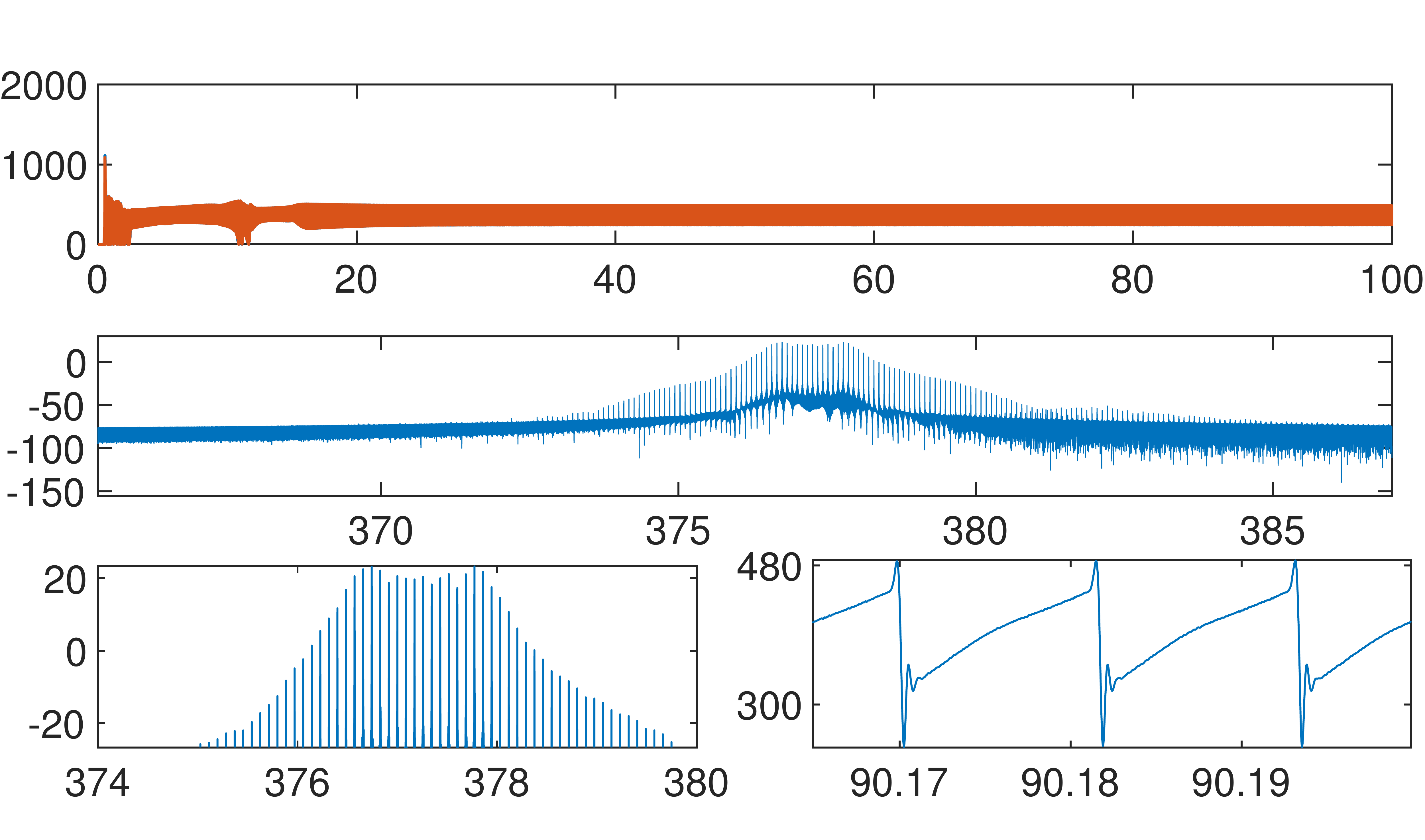}
\put(-3,48){{$P$}}
\put(-6,29){{$|\hat{E}|^2$}}
\put(-6,14){{$|\hat{E}|^2$}}
\put(51,14){{$P$}}
\end{overpic}}
\caption{Evolution of output power $P$ (mW) and power spectral density $|\hat{E}|^2$ (dBm/Hz) of the full propagating model evaluated with the predictor corrector scheme. (a) (b) The temporal output and the power spectral density of the temporal output in log scale of the first waveguide at $I_{1in}$=100 mA with the coupling factor C=1. (c) (d) The zoomed power spectral density and temporal output of the first waveguide. The predictor corrector scheme shows robust stability numerically with $\Delta t=30fs$.}
\label{C1_full}
\end{figure}

\noindent Here the dimensionless coupling factor $C$ is determined by the design parameters and spacing of the waveguide array. Thus it can be easily adjusted via designing the waveguide array to realize optimal mode-locking of the output. 

Our simulations applied the numerical methods developed previously to the equation set above. 
With both propagating and counter propagating waves in the cavity, as well as the existence of chromatic dispersion and gain, we consider using the predictor-corrector scheme with transmission and reflection at the boundary and propagate in the direction of time. For our simulation, we use material parameters of GaAs for a higher central transition energy. The chromatic dispersion coefficient is 1.25 $ps^2/m$, and the full material parameters used in the simulation are listed in table \ref{material_param} in Appendix. The coupling factor $C=1$ is used between the waveguides in the array and the input current applied to the first waveguide is $100mA$. 

\begin{figure}[t]
{\begin{overpic}
[width=0.95\linewidth]{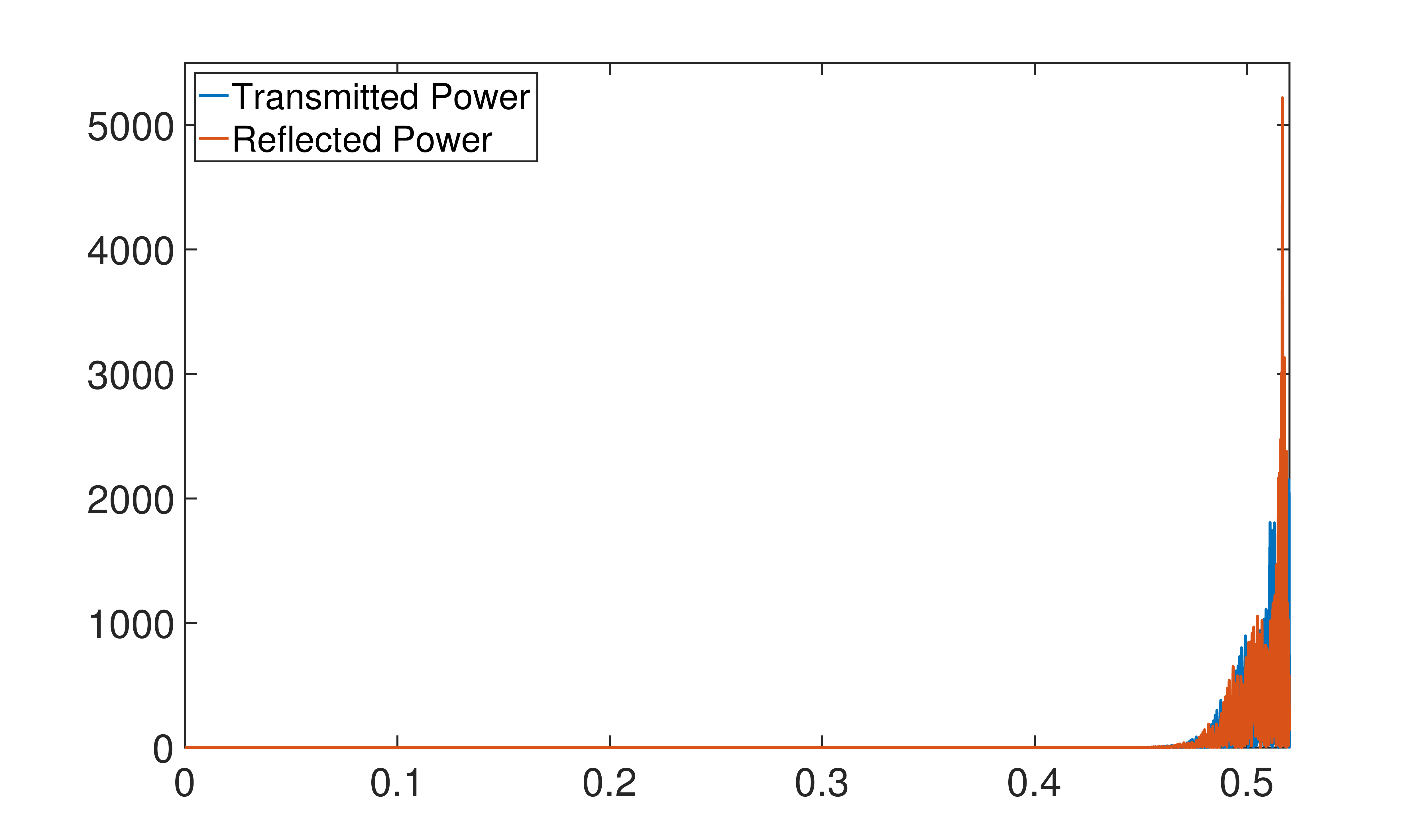}
\put(2,35){{$P$}}
\end{overpic}}
\caption{Evolution of output power $P$ (mW) of the full propagating model in the first waveguide evaluated with the predictor corrector scheme. Discretized time step is set to be $\Delta t=50fs$ while all the other parameters maintain the same as previously. The numerical solution becomes unstable with an increased time step size and eventually explodes at around $t=0.5ns$.}
\label{dt50}
\end{figure}

To numerically demonstrate the stability of the predictor corrector scheme developed in Section 4B, the waveguide array governing equations is simulated over a large number of round trips. The propagating and counter propagating waves in the first waveguide are stably evolved as shown in Fig. ~\ref{C1_full}. Here we propagated numerically along the characteristic, that is, each time and spatial step are discretized so that $\Delta z/\Delta t=v_g$, and $\Delta t$ is chosen to be $30fs$. The output power and spectral density show a repeatable waveform and broad spectral lines consistent with a mode-locked state over a long simulation time.  Moreover,  the predictor-corrector scheme shows robust stability compared to other methods with both wave propagation and chromatic dispersion included.   It can also be unstable if the time step is taken to be too large as shown in Fig. ~\ref{dt50}, where the time step size is increased from $30fs$ to $50fs$ in the simulation. The predictor corrector scheme is unstable with increased stepping size and generates numerical blowup quickly at around $t=0.5ns$. 

\section{Conclusion}
In conclusion, we have explored the stability of a number of numerical time-stepping schemes including Euler stepping, backward Euler, leap frog (2,2), predictor-corrector, and two new numerical schemes that integrate predictor-corrector and operator-splitting strategies.  Given the diversity of physics for our models of interest, including nonlinear dispersive wave equations with counter propagating waves and complex gain dynamics (semiconductor physics), these new schemes provide viable strategies that are not only stable, but computationally efficient when evaluating the semiconductor physics.    In general, the predictor corrector scheme shows reliable stability with a properly chosen step size. However the introduction of the operator splitting technique is motivated by the computational of the gain.  Indeed, the combination of the operator-splitting scheme with the predictor-corrector is capable of efficiently decreasing the time complexity for evaluating the complex gain term in the predictor step, consequently providing an excellent numerical strategy when considering both stability and efficiency.

We have numerically validated the reliability of the predictor-corrector scheme with operator-splitting by applying it to a comprehensive traveling wave model for a quantum well and the mode-coupling in a waveguide array. The propagating and counter propagating waves are stably evolved by the predictor corrector scheme with time stepping size near $\Delta t=30fs$. For larger step sizes, the stability of the predictor-corrector is compromised.  Thus much like a CFL requirement, our proposed method requires a sufficiently small  stepping size to balance to ensure stability of the scheme.   Regardless, the stable numerical results characterize the generation of frequency combs in the coupled waveguide array, diode laser.  The predictor-corrector scheme along with the operator-splitting can be broadly applied, including to more complicated wave propagation models.

\section*{Acknowledgments}

J. N. Kutz acknowledges support from the Air Force Office of Scientific Research (FA9550-17-1-0200).
This research was also developed in part with funding from the Defense Advanced Research Projects Agency (DARPA) through the SCOUT program. The views, opinions and/or findings expressed are those of the author and should not be interpreted as representing the official views or policies of the Department of Defense or the U.S. Government. This research was also supported in part through computational resources and services provided by Advanced Research Computing at the University of Michigan, Ann Arbor.

\section*{}
Approved for Public Release; Distribution Unlimited. Public Release Case Number 19-0457.

\section* {Appendix}
\subsection{Derivation of the total polarization}
The material polarization $P_{tot}$ is obtained from the Bloch equations as tailored to semiconductors \cite{ChowKochSargent1994}:
\begin{subequations}
\begin{align}
\begin{split}
\label{sc_bloch_p}
i \hbar \pd{p(\v{k},t)}{t} = (\hbar\omega_0 - \Delta E_{cv}(\v{k}) )p(\v{k},t) \\
- \frac{d_{cv} }{2}E(\v{k}, z,t)(\rho^e(\v{k},t)+ \rho^h(\v{k},t) - 1)-i\hbar \frac{p(\v{k}, t)}{T_2}
\end{split}
\end{align}
\begin{align}
\label{sc_bloch_ne}
\pd{\rho^e(\v{k},t)}{t} &= -\frac{1}{\hbar} Im[d^*_{cv} E^*(z,t) p(\v{k},t)] + \pd{\rho^e(\v{k},t)}{t}|_{relax}\\
\label{sc_bloch_nh}
\pd{\rho^h(\v{k},t)}{t} &= -\frac{1}{\hbar} Im[d^*_{cv} E^*(z,t) p(\v{k},t)] + \pd{\rho^h(\v{k},t)}{t}|_{relax}
\end{align}
\end{subequations}
where $p(\v{k},t)$ is the microscopic polarization, $\rho^{e,h}(\bf{k},t)$ is the occupation probability of electrons and holes, $d_{cv}$ is the dipole matrix element, $\Delta E_{cv}(\bf{k})$ is the transition energy between the conduction and valence bands, and $T_2 = 1/\Gamma$ is the intraband relaxation time which gives rise to homogenous broadening. It is important to note that these equations are in the time domain but are parameterized by the wavevector $\bf{k}$ and hence represent the time evolution of the subset of carriers with momentum $\bf{k}$.

A key simplification in our model is to assume that the intraband scattering is sufficiently fast to warrant the microscopic polarization adiabatically following the changes in carrier population. For modeling ultra-short pulses, this assumption may no longer hold and a full set of polarization equations will need to be solved dynamically. Integrating Eq.~(\ref{sc_bloch_p}), we obtain a time domain expression for the microscopic polarization in terms of the occupation probabilities and the electric field:

\begin{align}
\begin{split}
p(\v{k}, t) = \frac{i d_{cv} E(\v{k},z,t)}{2\hbar} \int_{-\infty}^t dt' E(z,t') e^{ -\left(i\frac{\Delta E_{cv}(\v{k})}{\hbar}-\omega_0 \right)(t-t')-\Gamma(t-t')} \\
\times (\rho^e(\v{k},t') + \rho^h(\v{k},t') - 1)
\end{split}
\end{align}
Next, we make the standard adiabatic approximation in which we assume the occupation probabilities evolve slowly compared to the intraband relaxation time $1/\Gamma$ and can be taken out of the integral, with $t'$ replaced by $t$. The remaining convolution integral is then defined as the filtered field \cite{Gioannini2015}

\begin{align}
\label{F_field}
F(\v{k}, z,t) = \Gamma \int_{-\infty}^t dt' e^{i(\frac{\Delta E_{cv}(\v{k}) }{\hbar}-\omega_0)(t-t')-\Gamma(t-t')} E(z,t')
\end{align}

The filtered field consists of all the components that interact with the population $\rho^{e,h}(\bf{k},t)$. Here the transition frequency is defined such that $\hbar\omega_0$ is the transition energy for a confined electron-hole pair with zero transverse energy and satisfies

\begin{align*}
\frac{\Delta E_{cv}(\v{k}) }{\hbar}-\omega_0 = \frac{E_t(\v{k})}{\hbar}
\end{align*}
Thus each discretized carrier group will have a different filtering frequency defined by the transverse energy $E_t$. The time-dependent microscopic polarization reduces to a simple expression:

\begin{align}
\label{p_ss}
p(\v{k}, t) = \frac{i d_{cv}}{2\hbar \Gamma} F(\v{k},z,t)
\end{align}
Here we note that physically, the $\bf{k}$ dependence of the confined carriers in the quantum well is due to a momentum $\bf{k}$ in the two transverse directions, and we therefore define a transverse energy with a simple parabolic band structure:

\begin{align}
E_t &= \frac{\hbar^2 |\v{k}|^2}{2 m_r^*}
\end{align}

\noindent where $m^*_r$ is the reduced effective mass. Hence to save space, we interchangeably write $\rho^{e,h}(\v{k},t) \leftrightarrow \rho^{e,h}_{E_t}$. We can also rewrite the filtered field by interchanging $F(\v{k},z,t) \leftrightarrow F(E_t,z,t)$.

The total polarization per volume is a summation over all carrier groups with momentum $\bf{k}$. Therefore, the total polarization for a 2-D quantum well can be written as:

\begin{align}
\label{ptot_sumk}
P_{tot}(t) = \frac{2}{V} \sum_{\v{k}} d^*_{ev} p(\v{k},t) = i\frac{|d_{cv}|^2}{2\hbar\Gamma} \frac{2}{V} \sum_{\v{k}} (\rho^e_{E_t}+\rho^h_{E_t}-1) F(E_t,z,t).
\end{align}

\noindent The $\v{k}$-summation can be converted to a transverse energy integral. We use a simple parabolic dispersion relation for the conduction and valence bands:

\begin{subequations}
\begin{align}
E_c &= E_g+ E_{e1}+\frac{\hbar^2 |\v{k}|^2}{2 m_e^*}\\
E_v &= E_{h1}-\frac{\hbar^2 |\v{k}|^2}{2 m_h^*}\\
\hbar\omega_0 &= E_g+E_{e1} -E_{h1}
\end{align}
\end{subequations}

\noindent where $E_g$ is the band gap energy, $E_{e1}$ is the confined electron energy, $E_{h1}$ is the confined hole energy, $m^*_{e,h}$ is the electron and hole effective mass (we have assumed only a single confined electron state). Rewriting Eq.~(\ref{ptot_sumk}) with an energy integral, we obtain:

\begin{align}
\label{ptot_t}
P_{tot}(t) =  i\frac{|d_{cv}|^2}{2\hbar\Gamma} \int dE_t D_r^{2D} (\rho^e_{E_t}+\rho^h_{E_t}-1) F(E_t,z,t)
\end{align}

\subsection{Carrier rate equations for the SCH and QW sections}
The QW equations are labeled with the transverse variable  for each discretized bin yielding

\begin{subequations}
\begin{align}
\begin{split}
\label{sch_eq}
\pd{\rho^{e,h}_{sch}}{t} = \frac{\eta J_{in}}{q N_{c,v,sch}h_{sch}}(1-\rho^{e,h}_{sch})- \frac{\rho^{e,h}_{sch}}{\tau_{sp}} \\
+n_{qw}\sum_{E_t} \left[\rho^{e,h}_{qw, E_t}\frac{(1-\rho^{e,h}_{sch})}{\tau^{e,h}_e}-\rho^{e,h}_{sch}\frac{(1-\rho^{e,h}_{qw,E_t})}{\tau^{e,h}_c}\right]
\end{split}
\end{align}
\begin{align}
\begin{split}
\label{qw_eq}
\pd{\rho^{e,h}_{qw, E_t}}{t} = \frac{h_{sch}N_{c,v,sch}}{n_{qw} h_{qw}N_{r,qw}}\left(\rho^{e,h}_{sch}\frac{(1-\rho^{e,h}_{qw,E_t})}{\tau^{e,h}_c} - \rho^{e,h}_{qw,E_t}\frac{(1-\rho^{e,h}_{sch})}{\tau^{e,h}_e}\right) \\
- \frac{\rho^{e,h}_{qw, E_t}}{\tau_{sp}} - R_{st} - R_{g}
\end{split}
\end{align}
\begin{align}
\begin{split}
\label{pg_eq}
\pd{\rho_{g, E_t}}{t} &=- \frac{\rho_{g,E_t}}{\tau_{sp}} -4k_0^2 D \rho_{g,E_t} - 2g_0 \frac{\Delta E_t}{(\hbar\omega_0)^2 h_{qw} W N_{r,qw}}\\
& \hspace*{-.2in}\times \! \left[ \frac{1}{2}(E_+^*F_- \!+\! F_+^* E_-)(\rho^e_{qw}\!+\!\rho^h_{qw} \!-\!1) \!+\! 2\text{Re}(E^*_+ \!F_+ \!+\! E^*_-F_-)\rho_{g,E_t}\right]
\end{split}
\end{align}
\end{subequations}

\begin{align}
R_{st} &= 2g_0 \frac{\Delta E_t}{(\hbar\omega_0)^2 h_{qw} W N_{r,qw}} (\rho_{qw, E_t}^e+\rho^h_{qw, E_t}-1) \text{Re}(E^* F)
\end{align}

\begin{align}
\begin{split}
R_g &= 2g_0 \frac{\Delta E_t}{(\hbar\omega_0)^2 h_{qw} W N_{r,qw}} \big((E_+F^*_- + F_+ E^*_-)\rho_{g,E_t} \\
& + (E_+^*F_- + F_+^* E_-)\rho^*_{g,E_t}\big)
\end{split}
\end{align}

\noindent where $N_{c,v,sch} = 2 \left(\frac{m_{e,h}^* k_B T}{2\hbar^2 \pi}\right)^{3/2}$, $N_r = \frac{m_{r}^* \Delta E_t }{\hbar^2 \pi h_{qw}}$ are the effective 3-D and 2-D density of states, $D$ is the ambipolar diffusion coefficient, $\tau_{sp}$ is the spontaneous emission lifetime, $\tau^{e,h}_c$ is the capture lifetime, and $\tau^{e,h}_e$ is the escape lifetime. The recombination rates $R_{st}$ and $R_g$ govern population decay due to stimulated emission and the carrier grating respectively. The escape times $\tau_e^{e,h}$ are particularly important in our model as they phenomenologically represent intraband interactions.  As shown in the Appendix, they are given by 

\begin{align}
\tau^e_e = \tau^e_c \exp( (\delta E_c-\frac{m^*_r}{m^*_e}E_t)/k_B T)\\
\tau^h_e = \tau^h_c \exp( (\delta E_v-\frac{m^*_r}{m^*_h}E_t)/k_B T)
\end{align}
The value of these escape times is tailored specifically to allow the rate equations \ref{sch_eq}, \ref{qw_eq} to relax to the Fermi-Dirac distribution.

\subsection{Simulation parameters for the GaAs system}

Table~\ref{ta:para} details the parameters used in simulations presented in Sec.~5.  

\begin{table}[t]
\centering
\caption{\bf Simulation parameters for the GaAs system. \label{ta:para}}
\begin{tabular}{lllllll}
\hline
{\small Symbol} & Description & Value\\ \hline
$L$ & Length of device & 500 $\mu$m \\ \hline
$W$ & Width of waveguide & 4 $\mu$m \\ \hline
$h_{sch}$ & Height of SCH layer & 50 nm \\ \hline
$h_{qw}$ & Height of quantum well & 5 nm \\ \hline
$n_0$ & Group refractive index & 3.5 \\ \hline
$n_{qw}$ & Number of quantum wells & 2 \\ \hline
$\alpha$ & Intrinsic waveguide loss & 5 cm$^{-1}$ \\ \hline
$\Gamma_{xy}$ & Optical confinement factor & 0.02 \\ \hline
$\alpha_S$ & Two-photon absorption & 580 W$^{-1}$m$^{-1}$\\ \hline
$\beta_S$ & Kerr coefficient &  430 W$^{-1}$m$^{-1}$\\ \hline
$\hbar \omega_0$ & Central transition energy & 1.55 eV\\ \hline
$|\uv{e}\cdot\v{p}|^2 $ & Momentum matrix element & 25 meV$\!\times \! m_0/6$ \\ \hline
$\Gamma$ & Homogenous half linewidth & 11 meV/$\hbar$ \\ \hline
$m^*_{e, sch}$ & {\small Effective mass of electrons in SCH layer} & $0.125 m_0$ \\ \hline
$m^*_{h, sch}$ &  {\small Effective mass of holes in SCH layer} & $0.703 m_0$ \\ \hline
$m^*_{e, qw}$ &  {\small Effective mass of electrons in GaAs QW} & $0.093 m_0$\\ \hline
$m^*_{h, qw}$ &  {\small Effective mass of holes in GaAs QW} & $0.53 m_0$ \\ \hline
$\tau_c^{e, h,qw}$ & electron, hole capture time  & $1$, 10 ps \\ \hline
$\delta E_c$ &  {\small Conduction band quantum well barrier} & 50 meV\\ \hline
$\delta E_v$ & Valence band quantum well barrier &  25 meV\\ \hline
$\beta_{sp}$ &  {\small Spontaneous emission coupling factor}& $1\times 10^{-4}$ \\ \hline
$\tau_{sp}$ & Spontaneous emission lifetime & 1 ns \\ \hline
$D$ & Ambipolar diffusion coefficient & 20 cm$^2$/s \\ \hline
\end{tabular}
  \label{material_param}
\end{table}

\bibliography{refs,QWLaser_Master}

\providecommand{\noopsort}[1]{}\providecommand{\singleletter}[1]{#1}%
\begin{thebibliography}{10}
\newcommand{\enquote}[1]{``#1''}

\bibitem{twin}
B.~Schleich, N.~Anwer, L.~Mathieu, and S.~Wartzack, \enquote{Shaping the
  digital twin for design and production engineering,} CIRP Annals \textbf{66},
  141--144 (2017).

\bibitem{MarkIEEE}
M.~Dong, N.~M. Mangan, J.~N. Kutz, S.~T. Cundiff, and H.~G. Winful,
  \enquote{Traveling wave model for frequency comb generation in single-section
  quantum well diode lasers,} IEEE Journal of Quantum Electronics \textbf{53},
  1--11 (2017).

\bibitem{Moskalenko2017}
V.~Moskalenko, J.~Koelemeij, K.~Williams, and E.~Bente, \enquote{Study of extra
  wide coherent optical combs generated by a qw-based integrated passively
  mode-locked ring laser,} Opt. Letters \textbf{42}, 1428 (2017).

\bibitem{Rosales2011}
R.~Rosales, K.~Merghem, A.~Martinez, A.~Akrout, J.-P. Tourrenc, A.~Accard,
  F.~Lelarge, and A.~Ramdane, \enquote{Inas/inp quantum-dot passively
  mode-locked lasers for 1.55-$\mu$m applications,} IEEE J. Sel. Top. Quantum
  Electron. \textbf{17}, 1292 (2011).

\bibitem{Delfyett1992}
P.~J. Delfyett, L.~T. Florez, N.~Stoffel, T.~Gmitter, N.~C. Andreadakis,
  Y.~Silberberg, and J.~P. Heritage, \enquote{High-power ultrafast laser
  diodes,} IEEE J. Quantum Electron. \textbf{28}, 2203 (1992).

\bibitem{Tiemeijer1989}
L.~F. Tiemeijer, P.~I. Kuindersma, P.~J.~A. Thijs, and G.~L.~J. Rikken,
  \enquote{Passive fm locking in ingaasp semiconductor lasers,} IEEE J. Quantum
  Electron. \textbf{25}, 1385 (1989).

\bibitem{Gioannini2015}
M.~Gioannini, P.~Bardella, and I.~Montrosset, \enquote{Time-domain
  traveling-wave analysis of the multimode dynamics of quantum dot fabry--perot
  lasers,} IEEE Sel. Topics Quantum Electron. \textbf{21}, 1900811 (2015).

\bibitem{Rosales2012}
R.~Rosales, K.~Merghem, C.~Calo, G.~Bouwmans, I.~Krestnikov, A.~Martinez, and
  A.~Ramdane, \enquote{Optical pulse generation in single section inas/gaas
  quantum dot edge emitting lasers under continuous wave operation,} App. Phys.
  Lett. \textbf{101}, 221113 (2012).

\bibitem{Rosales2012-2}
R.~Rosales, S.~G. Murdoch, R.~Watts, K.~Merghem, A.~Martinez, F.~Lelarge,
  A.~Accard, L.~P. Barry, and A.~Ramdane, \enquote{High performance mode
  locking characteristics of single section quantum dash lasers,} Optics
  Express \textbf{20}, 8649 (2012).

\bibitem{Sato2003}
K.~Sato, \enquote{Optical pulse generation using fabry--p{\'e}rot lasers under
  continuous-wave operation,} IEEE J. Sel. Top. Quantum Electron. \textbf{9},
  1288 (2003).

\bibitem{Calo2015}
C.~Cal{\`o}, V.~Vujicic, R.~Watts, C.~Browning, K.~Merghem, V.~Panapakkam,
  F.~Lelarge, A.~Martinez, B.-E. Benkelfat, A.~Ramdane, and L.~P. Barry,
  \enquote{Single-section quantum well mode-locked laser for 400 gb/s ssb-ofdm
  transmission,} Opt. Express \textbf{23}, 26442 (2015).

\bibitem{Homar1996}
M.~Homar, S.~Balle, and M.~S. Miguel, \enquote{Mode competition in a
  fabry-perot semiconductor laser: travelling wave model with asymmetric
  dynamical gain,} Optics Communications \textbf{131}, 380--390 (1996).

\bibitem{Arakawa1986}
Y.~Arakawa and A.~Yariv, \enquote{Quantum well lasers - gain, spectra,
  dynamics,} IEEE J. Quantum Electron. \textbf{QE22}, 1887 (1986).

\bibitem{KN2010}
S.~N. Kaunga-Nyirenda, M.~P. Dlubek, A.~J. Phillips, J.~J. Lim, E.~C. Larkins,
  and S.~Sujecki, \enquote{Theoretical investigation of the role of optically
  induced carrier pulsations in wave mixing in semiconductor optical
  amplifiers,} J. Opt. Soc. Am. B \textbf{27}, 168 (2010).

\bibitem{McDonald1995}
D.~McDonald and R.~F. O'Dowd, \enquote{Comparison of two- and three-level rate
  equations in the modeling of quantum-well lasers,} IEEE J. Quantum Electron.
  \textbf{31}, 1927 (1995).

\bibitem{Jones1995}
D.~J. Jones, L.~M. Zhang, J.~E. Carroll, and D.~D. Marcenac, \enquote{Dynamics
  of monolithic passively mode-locked semiconductor lasers,} IEEE J. Quantum
  Electron. \textbf{31}, 1051--1058 (1995).

\bibitem{Vandermeer2005}
A.~D. Vandermeer and D.~T. Cassidy, \enquote{A rate equation model of
  asymmetric multiple quantum-well lasers,} IEEE J. Quantum Electron.
  \textbf{41}, 917 (2005).

\bibitem{Gordon2008}
A.~Gordon, C.~Y. Wang, L.~Diehl, F.~X. K{\"a}rtner, A.~Belyanin, D.~Bour,
  S.~Corzine, G.~H{\"o}fler, H.~C. Liu, H.~Schneider, T.~Maier, M.~Troccoli,
  J.~Faist, and F.~Capasso, \enquote{Multimode regimes in quantum cascade
  lasers: From coherent instabilities to spatial hole burning,} Phys. Rev. A
  \textbf{77}, 053804 (2008).

\bibitem{Lenstra2014}
D.~Lenstra and M.~Yousefi, \enquote{Rate-equation model for multi-mode
  semiconductor lasers with spatial hole burning,} Opt. Express \textbf{22},
  8144 (2014).

\bibitem{Chow2002}
W.~W. Chow, H.~C. Schneider, S.~W. Koch, C.-H. Chang, L.~Chrostowski, and C.~J.
  Chang-Hasnain, \enquote{Nonequilibrium model for semiconductor laser
  modulation response,} IEEE J. Quantum Electron. \textbf{38}, 402 (2002).

\bibitem{Kutz:2013}
J.~N. Kutz, \emph{Data-Driven Modeling \& Scientific Computation: Methods for
  Complex Systems \& Big Data} (Oxford University Press, 2013).

\bibitem{hudson1}
D.~D. Hudson, K.~Shish, T.~R. Schibli, J.~N. Kutz, D.~N. Christodoulides,
  R.~Morandotti, and S.~T. Cundiff, \enquote{Nonlinear femtosecond pulse
  reshaping in waveguide arrays,} Optics letters \textbf{33}, 1440--1442
  (2008).

\bibitem{hudson2}
D.~D. Hudson, J.~N. Kutz, T.~R. Schibli, D.~N. Christodoulides, R.~Morandotti,
  and S.~T. Cundiff, \enquote{Spatial distribution clamping of discrete spatial
  solitons due to three photon absorption in algaas waveguide arrays,} Optics
  express \textbf{20}, 1939--1944 (2012).

\bibitem{Javaloyes2009}
J.~Javaloyes and S.~Balle, \enquote{Emission directionality of semiconductor
  ring lasers: A traveling-wave description,} IEEE J. Quantum Electron.
  \textbf{45}, 431 (2009).

\bibitem{Homar1996-2}
M.~Homar, J.~V. Moloney, and M.~S. Miguel, \enquote{Traveling wave model of a
  multimode fabry-perot laser in free running and external cavity
  configurations,} IEEE J. Quantum Electron. \textbf{32}, 553 (1996).

\bibitem{WilliamsWGA}
X.~Zhang, M.~Williams, S.~T. Cundiff, and J.~N. Kutz, \enquote{Semiconductor
  diode laser mode-locking by a waveguide array,} IEEE Journal of Selected
  Topics in Quantum Electronics \textbf{22}, 34--39 (2016).

\bibitem{proctor05}
J.~L. Proctor and J.~N. Kutz, \enquote{Passive mode-locking by use of waveguide
  arrays,} Optics letters \textbf{30}, 2013--2015 (2005).

\bibitem{kutz08}
J.~N. Kutz and B.~Sandstede, \enquote{Theory of passive harmonic mode-locking
  using waveguide arrays,} Optics Express \textbf{16}, 636--650 (2008).

\bibitem{ching12}
Q.~Chao, D.~D. Hudson, J.~N. Kutz, and S.~Cundiff, \enquote{Waveguide array
  fiber laser,} IEEE Photonics Journal \textbf{4}, 1438--1442 (2012).

\bibitem{ChowKochSargent1994}
W.~W. Chow, S.~W. Koch, and M.~S. III, \emph{Semiconductor-Laser Physics}
  (Springer-Verlag, 1994).

\end{thebibliography}


\end{document}